\documentclass[12pt]{article}
\usepackage{graphicx}
\usepackage{amsmath}
\usepackage{amssymb}
\usepackage{authblk}

\begin{document}

\title{Possible measurable effects of light propagating in electromagnetized vacuum, as predicted by a scalar tensor theory of gravitation}

\author{T. E. Raptis$^1$ and F. O. Minotti$^2$}
\affil{$^1$Division of Applied Technologies\\National Center for Science and Research "Demokritos", Athens, Greece.\\$^2$Departamento de F\'{\i}sica, Facultad de Ciencias Exactas and Naturales,Universidad de Buenos Aires \\ Instituto de F\'{\i}sica del Plasma (CONICET), Buenos Aires, Argentina}

\maketitle
\footnote[1]{rtheo@dat.demokritos.gr}
\footnote[2]{minotti@df.uba.ar}

\begin{abstract}
The effect of static electromagnetic fields on the propagation of light is analyzed in the context of a particular class of scalar-tensor gravitational theories. It is found that for appropriate field configurations and light polarization, anomalous amplitude variations of the light as it propagates in either a magnetized or electrified vacuum are strong enough to be detectable in relatively simple laboratory experiments.
\end{abstract}

\section{Introduction}

Scalar-tensor (ST) gravitational theories are the most firm candidates for
extensions of General Relativity (GR). A great part of their interest comes
from the fact that they are induced naturally in the reduction to four
dimensions of string and Kaluza-Klein models\cite{fujiibook,chavineau},
resulting mostly in the form of a Brans-Dicke (BD) type of ST theory\cite%
{bransdicke}, often involving also non-minimal coupling to matter, leading
to the so called fifth force\cite{fifthforce}. It is also interesting that
ST theories are shown to be mathematically equivalent to theories with
action depending non-linearly on the Ricci scalar, the so called $f\left(
R\right) $ theories\cite{frt}. Finally, ST theories are possibly the
simplest extension of GR that could accommodate cosmological issues as
inflation and universe-expansion acceleration, as well as possible
space-time variation of fundamental constants\cite{bekenstein}. On the other
hand, observational and experimental evidence puts strong limits to the
observable effects of a possible scalar field. For example, in the case of a
massless scalar the BD parameter $\omega $ is constrained by precise
Solar-System experiments to be a large number ($\omega >4\times 10^{4}$)\cite%
{bertotti}. In this way, ST gravity phenomenology appears to be very similar
to that of GR, thus putting strong limits to possible experimental verifications. There is however a very interesting extension of ST gravity
put forward by Mbelek and Lachi\`{e}ze-Rey\cite{MLR}, which could allow
electromagnetic (EM) fields to modify the space-time metric far more
strongly than predicted by GR and standard ST theories. The theory was
applied in cosmological\cite{mbelek2003} and galactic\cite{mbelek2004}
contexts, and in\cite{MLR} it was used to explain the discordancy in the
measurements of Newton gravitational constant as due to the effect of the
Earth's magnetic field. The key new element of that theory is an additional,
external scalar field $\psi $, minimally coupled to gravity. In\cite{minotti}
it was shown that a general ST theory that includes an external field $\psi $
with the mentioned characteristics, and with the magnitude of the coupling
derived in\cite{MLR}, can explain the unusual forces on asymmetric resonant
cavities recently reported\cite{juan}.

\section{Scalar-tensor theory}

We will consider the weak-field limit of a ST theory with action given by
(SI units are used) 
\begin{eqnarray}
S &=&-\frac{c^{3}}{16\pi G_{0}}\int \sqrt{-g}\phi Rd\Omega +\frac{c^{3}}{%
16\pi G_{0}}\int \sqrt{-g}\frac{\omega \left( \phi \right) }{\phi }\nabla
^{\nu }\phi \nabla _{\nu }\phi d\Omega  \notag \\
&&+\frac{c^{3}}{16\pi G_{0}}\int \sqrt{-g}\phi \left[ \frac{1}{2}\nabla
^{\nu }\psi \nabla _{\nu }\psi -U\left( \psi \right) -J\psi \right] d\Omega 
\notag \\
&&-\frac{\varepsilon _{0}c}{4}\int \sqrt{-g}\lambda \left( \phi \right)
F_{\mu \nu }F^{\mu \nu }d\Omega -\frac{1}{c}\int \sqrt{-g}j^{\nu }A_{\nu
}d\Omega  \notag \\
&&+\frac{1}{c}\int \mathcal{L}_{mat}d\Omega .  \label{SKK}
\end{eqnarray}
In (\ref{SKK}) the internal, non-dimensional scalar field is $\phi $, while
the external scalar field is $\psi $. These fields have vacuum expectation
values (VEV) $\phi _{0}=1$ and $\psi _{0}$, respectively. $G_{0}$ represents
Newton gravitational constant, $c$ is the velocity of light in vacuum, and $%
\varepsilon _{0}$ is the vacuum permittivity. $\mathcal{L}_{mat}$ is the
lagrangian density of matter. The other symbols are also conventional, $R$
is the Ricci scalar, and $g$ the determinant of the metric tensor $g_{\mu
\nu }$. The BD parameter $\omega \left( \phi \right) $ is considered a
function of $\phi $, as it usually results in the reduction to four
dimensions of multidimensional theories\cite{chavineau}. The function $%
\lambda \left( \phi \right) $ in the term of the action of the EM field is
of the type appearing in Bekenstein's theory and other effective theories%
\cite{mbelek2003}. The EM tensor is $F_{\mu \nu }=\nabla _{\mu }A_{\nu
}-\nabla _{\nu }A_{\mu }$, \ given in terms of the EM quadri-vector $A_{\nu
} $, with sources given by the quadri-current $j^{\nu }$. $U$ and $J$ are,
respectively, the potential and source of the field $\psi $. The source $J$
contains contributions from the matter, EM field and the scalar $\phi $. The
model for $J$ proposed in\cite{MLR} is 
\begin{equation}
J=\beta _{mat}\left( \psi ,\phi \right) \frac{8\pi G_{0}}{c^{4}}%
T^{mat}+\beta _{EM}\left( \psi ,\phi \right) \frac{4\pi G_{0}\varepsilon _{0}%
}{c^{2}}F_{\mu \nu }F^{\mu \nu },  \label{source}
\end{equation}%
where $T^{mat}$ is the trace of the energy-momentum tensor of matter,\ 
\begin{equation*}
T_{\mu \nu }^{mat}=-\frac{2}{\sqrt{-g}}\frac{\delta \mathcal{L}_{mat}}{%
\delta g^{\mu \nu }}.
\end{equation*}

Variation of (\ref{SKK}) with respect to $g^{\mu \nu }$ results in ($T_{\mu
\nu }^{EM}$ is the usual electromagnetic energy tensor, and $T_{\mu \nu
}^{\phi }$ the energy tensor associated to the scalar $\phi $)%
\begin{eqnarray}
\phi \left( R_{\mu \nu }-\frac{1}{2}Rg_{\mu \nu }\right) &=&\frac{8\pi G_{0}%
}{c^{4}}\left[ \lambda \left( \phi \right) T_{\mu \nu }^{EM}+T_{\mu \nu
}^{mat}\right] +T_{\mu \nu }^{\phi }  \notag \\
&&+\frac{\phi }{2}\left( \nabla _{\mu }\psi \nabla _{\nu }\psi -\frac{1}{2}%
\nabla ^{\gamma }\psi \nabla _{\gamma }\psi g_{\mu \nu }\right)  \notag \\
&&+\frac{\phi }{2}\left( U+J\psi \right) g_{\mu \nu }.  \label{Glm}
\end{eqnarray}

Variation with respect to $\phi $ gives 
\begin{eqnarray*}
\phi R+2\omega \nabla ^{\nu }\nabla _{\nu }\phi &=&\left( \frac{\omega }{%
\phi }-\frac{d\omega }{d\phi }\right) \nabla ^{\nu }\phi \nabla _{\nu }\phi -%
\frac{4\pi G_{0}\varepsilon _{0}}{c^{2}}\phi \frac{d\lambda }{d\phi }F_{\mu
\nu }F^{\mu \nu } \\
&&-\frac{\partial J}{\partial \phi }\psi \phi +\phi \left[ \frac{1}{2}\nabla
^{\nu }\psi \nabla _{\nu }\psi -U\left( \psi \right) -J\psi \right] ,
\end{eqnarray*}
which can be rewritten, using the contraction of (\ref{Glm}) with $g^{\mu
\nu }$ to replace $R$, as 
\begin{eqnarray}
\left( 2\omega +3\right) \nabla ^{\nu }\nabla _{\nu }\phi &=&-\frac{d\omega 
}{d\phi }\nabla ^{\nu }\phi \nabla _{\nu }\phi -\frac{4\pi G_{0}\varepsilon
_{0}}{c^{2}}\phi \frac{d\lambda }{d\phi }F_{\mu \nu }F^{\mu \nu }+\frac{8\pi
G_{0}}{c^{4}}T^{mat}  \notag \\
&&+\phi \left[ \frac{1}{2}\nabla ^{\nu }\psi \nabla _{\nu }\psi -U\left(
\psi \right) -J\psi \right] -\frac{\partial J}{\partial \phi }\psi \phi ,
\label{phi}
\end{eqnarray}
where it was used that $T^{EM}=T_{\mu \nu }^{EM}g^{\mu \nu }=0$.

The non-homogeneous Maxwell equations are obtained by varying (\ref{SKK})
with respect to $A_{\nu }$, 
\begin{equation}
\nabla _{\mu }\left\{ \lambda \left( \phi \right) F^{\mu \nu }\right\} =\mu
_{0}j^{\nu }.  \label{Maxwell}
\end{equation}%
with $\mu _{0}$ the vacuum permeability.

Finally, the variation with respect to $\psi $ results in 
\begin{equation}
\nabla ^{\nu }\nabla _{\nu }\psi +\frac{1}{\phi }\nabla ^{\nu }\psi \nabla
_{\nu }\phi =-\frac{\partial U}{\partial \psi }-J-\frac{\partial J}{\partial
\psi }\psi +\frac{\beta }{\phi }\frac{8\pi G_{0}}{c^{4}}T^{mat}.  \label{psi}
\end{equation}

Having included $G_{0}$, it is understood that $\phi $ takes values around
its vacuum expectation value (VEV) $\phi _{0}=1$. The scalar $\psi $ is also
dimensionless and of VEV $\psi _{0}$.

These equations can be approximated in the weak-field limit keeping only the
lowest significant order in the perturbations $h_{\mu \nu }$ of the metric $%
g_{\mu \nu }$ about the Minkowski metric $\eta _{\mu \nu }$, with signature
(1,-1,-1,-1), and of the scalar fields about their VEV $\phi _{0}$ and $\psi
_{0}$ 
\begin{equation}
-\eta ^{\gamma \delta }\partial _{\gamma \delta }\overline{h}_{\mu \nu
}=2\left( \partial _{\mu \nu }\phi -\eta ^{\gamma \delta }\partial _{\gamma
\delta }\phi \eta _{\mu \nu }\right) ,  \label{Gik0}
\end{equation}%
\begin{equation}
\partial _{\gamma }\overline{h}_{\nu }^{\gamma }=0,  \label{LG}
\end{equation}%
\begin{equation}
\left( 2\omega _{0}+3\right) \eta ^{\gamma \delta }\partial _{\gamma \delta
}\phi =-\left. \frac{\partial J}{\partial \phi }\right\vert _{\phi _{0},\psi
_{0}}\psi _{0},  \label{dphi0}
\end{equation}%
\begin{equation}
\partial _{\nu }F^{\mu \nu }=-\mu _{0}\left[ 1-\lambda _{0}^{\prime }\left(
\phi -\phi _{0}\right) \right] j^{\mu }-F^{\mu \nu }\partial _{\nu }\left(
\lambda _{0}^{\prime }\phi -\overline{h}/2\right)  \label{Max0}
\end{equation}%
\begin{equation}
\eta ^{\gamma \delta }\partial _{\gamma \delta }\psi =-\left. \frac{\partial
J}{\partial \psi }\right\vert _{\phi _{0},\psi _{0}}\psi _{0},  \label{dpsi0}
\end{equation}%
where $\omega _{0}=\omega \left( \phi _{0}\right) $, $\lambda _{0}^{\prime
}\equiv \left. d\lambda /d\phi \right\vert _{\phi _{0}}$, and 
\begin{equation*}
\overline{h}_{\mu \nu }\equiv h_{\mu \nu }-\frac{1}{2}h\eta _{\mu \nu },
\end{equation*}%
with $\overline{h}=\eta ^{\mu \nu }\overline{h}_{\mu \nu }$. In these
equations only the EM sources were included, the effect of the matter terms
either being negligible or included in the local gravitational field.

Using the source $J$ given in (\ref{source}), with only the EM terms we
finally obtain the complete set of equations for the EM field (making
explicit the electric and magnetic field vectors $\mathbf{E}$ and $\mathbf{B}
$, respectively) 
\begin{subequations}
\label{linearEM}
\begin{eqnarray}
\square \Theta &=&\varkappa \left( B^{2}-E^{2}/c^{2}\right) , \\
\mathbf{\nabla }\cdot \mathbf{E} &=&\frac{\widetilde{\rho }}{\varepsilon _{0}%
}-\mathbf{\nabla }\Theta \cdot \mathbf{E}, \\
\mathbf{\nabla }\times \mathbf{E} &=&-\frac{\partial \mathbf{B}}{\partial t}%
,\;\;\;\mathbf{\nabla }\cdot \mathbf{B}=0, \\
\mathbf{\nabla }\times \mathbf{B} &=&\mu _{0}\widetilde{\mathbf{j}}+\frac{1}{%
c^{2}}\frac{\partial \mathbf{E}}{\partial t}+\frac{1}{c^{2}}\frac{\partial
\Theta }{\partial t}\mathbf{E}-\mathbf{\nabla \Theta }\times \mathbf{B},
\end{eqnarray}%
where the auxiliary field $\Theta $ is defined as $\Theta \equiv \lambda
_{0}^{\prime }\phi -\overline{h}/2$, and the electromagnetic sources were
redefined as $\widetilde{j}^{\mu }\equiv \left[ 1-\lambda _{0}^{\prime
}\left( \phi -\phi _{0}\right) \right] j^{\mu }$. The constant $\varkappa $
is 
\end{subequations}
\begin{equation*}
\varkappa =-\frac{8\pi G_{0}\varepsilon _{0}\left( \lambda _{0}^{\prime
}-3\right) }{\left( 2\omega _{0}+3\right) c^{2}}\psi _{0}\left. \frac{%
\partial \beta _{EM}}{\partial \phi }\right\vert _{\phi _{0},\psi _{0}},
\end{equation*}%
which, according to\cite{MLR}, has a value of order%
\begin{equation}
\varkappa \simeq 5\times 10^{-8}\frac{A^{2}}{N^{2}}.  \label{chi}
\end{equation}

To study the propagation of electromagnetic waves we consider the case of a
vacuum with uniform and static electric and magnetic fields $\mathbf{E}_{0}$
and $\mathbf{B}_{0}$, so that one can linearize the system (\ref{linearEM})
in the perturbations as

\begin{subequations}
\label{full}
\begin{eqnarray}
\mathbf{\nabla }\cdot \delta \mathbf{E} &=&-\mathbf{\nabla \delta }\Theta
\cdot \mathbf{E}_{0}-\mathbf{\nabla }\Theta _{0}\cdot \delta \mathbf{E}, \\
\mathbf{\nabla }\times \delta \mathbf{E} &=&-\frac{\partial \delta \mathbf{B}%
}{\partial t},\;\;\;\mathbf{\nabla }\cdot \delta \mathbf{B}=0, \\
\mathbf{\nabla }\times \delta \mathbf{B} &=&\frac{1}{c^{2}}\frac{\partial
\delta \mathbf{E}}{\partial t}+\frac{1}{c^{2}}\frac{\partial \delta \Theta }{%
\partial t}\mathbf{E}_{0}-\mathbf{\nabla \delta \Theta }\times \mathbf{B}%
_{0}-\mathbf{\nabla \Theta }_{0}\times \delta \mathbf{B}, \\
\nabla ^{2}\Theta _{0} &=&-\varkappa \left( B_{0}^{2}-E_{0}^{2}/c^{2}\right)
, \\
\square \delta \Theta &=&2\varkappa \left( \mathbf{B}_{0}\cdot \delta 
\mathbf{B}-\mathbf{E}_{0}\cdot \delta \mathbf{E}/c^{2}\right) .
\end{eqnarray}

Starting with this system we consider now different simple
configurations.

\subsection{\protect\bigskip Case $\mathbf{E}_{0}=0$}

For the case without zero order electric field, $\mathbf{E}_{0}=0$, and
perturbations $\delta \mathbf{E}$, $\delta \mathbf{B}$, $\delta \Theta \sim
\exp i\left( \mathbf{k}\cdot \mathbf{x}-\omega t\right) $ one has from (\ref%
{full})

\end{subequations}
\begin{eqnarray*}
\mathbf{k}\cdot \delta \mathbf{E} &=&i\mathbf{\nabla }\Theta _{0}\cdot
\delta \mathbf{E}, \\
\mathbf{k}\times \delta \mathbf{E} &=&\omega \delta \mathbf{B},\;\; \\
\mathbf{k}\times \delta \mathbf{B} &=&-\frac{\omega }{c^{2}}\delta \mathbf{E}%
-\mathbf{k}\times \mathbf{B}_{0}\mathbf{\delta \Theta }+i\mathbf{\nabla
\Theta }_{0}\times \delta \mathbf{B}, \\
\left( k^{2}-\omega ^{2}/c^{2}\right) \delta \Theta &=&2\varkappa \mathbf{B}%
_{0}\cdot \delta \mathbf{B}.
\end{eqnarray*}

For propagation along the magnetic field, $\mathbf{B}_{0}\parallel \mathbf{k}
$, with $\mathbf{\nabla }\Theta _{0}$ $\perp \mathbf{k}$, one has

\begin{eqnarray*}
\delta E_{\shortparallel } &=&i\mathbf{\nabla }\Theta _{0}\cdot \delta 
\mathbf{E}_{\perp }/k, \\
\delta \mathbf{E}_{\perp } &=&-\frac{\omega }{k^{2}}\mathbf{k}\times \delta 
\mathbf{B},\;\; \\
\mathbf{k}\times \delta \mathbf{B} &=&-\frac{\omega }{c^{2}}\left( \delta 
\mathbf{E}_{\perp }+\delta E_{\shortparallel }\frac{\mathbf{k}}{k}\right) +i%
\mathbf{\nabla \Theta }_{0}\times \delta \mathbf{B},
\end{eqnarray*}%
replacement of the first two equations in the last results in%
\begin{equation*}
\left( 1-\frac{\omega ^{2}}{k^{2}c^{2}}\right) \mathbf{k}\times \delta 
\mathbf{B}+i\left[ \mathbf{\nabla \Theta }_{0}\times \delta \mathbf{B}-\frac{%
\omega ^{2}}{k^{4}c^{2}}\mathbf{\nabla }\Theta _{0}\cdot \left( \mathbf{k}%
\times \delta \mathbf{B}\right) \mathbf{k}\right] =0.
\end{equation*}

It is easy to see that the last equation can only be satisfied if 
\begin{equation*}
\omega ^{2}=k^{2}c^{2},
\end{equation*}%
the usual dispersion relation for EM waves in vacuum. For the kind of
propagation considered we thus have the standard plane EM wave for $\delta 
\mathbf{E}_{\perp }$ and $\delta \mathbf{B}$, with only the addition of a
longitudinal component of amplitude%
\begin{equation*}
\delta E_{\shortparallel }=i\mathbf{\nabla }\Theta _{0}\cdot \delta \mathbf{E%
}_{\perp }/k.
\end{equation*}

We consider now propagation perpendicular to the zero order field. Taking $%
\mathbf{E}_{0}=0$, $\mathbf{B}_{0}=B_{0}\mathbf{e}_{z}$, $\mathbf{\nabla }%
\Theta _{0}=a\mathbf{e}_{x}$, $\mathbf{k}=k_{x}\mathbf{e}_{x}+k_{y}\mathbf{e}%
_{y}$, the general system (\ref{full}) can be reduced to%
\begin{eqnarray*}
\left( \omega ^{2}-k^{2}c^{2}\right) \delta B_{z}+iak_{x}c^{2}\delta B_{z}
&=&k^{2}c^{2}B_{0}\delta \Theta , \\
\left( \omega ^{2}-k^{2}c^{2}\right) \delta B_{x}-iak_{y}c^{2}\delta B_{y}
&=&0, \\
\left( \omega ^{2}-k^{2}c^{2}\right) \delta B_{y}+iak_{x}c^{2}\delta B_{y}
&=&0, \\
\left( k^{2}c^{2}-\omega ^{2}\right) \delta \Theta &=&2\varkappa
c^{2}B_{0}\delta B_{z}.
\end{eqnarray*}

In the case $\delta B_{z}=0$, this system has non-trivial solution only for
the dispersion relation%
\begin{equation*}
\omega ^{2}-k^{2}c^{2}+iak_{x}c^{2}=0,
\end{equation*}%
from which, $\omega =kc+i\gamma $, with ($\cos \theta =k_{x}/k$) 
\begin{equation}
\gamma =-\frac{ac\cos \theta }{2}.  \label{small}
\end{equation}

In the case $\delta B_{z}\neq 0$, one also has $\omega =kc+i\gamma $, with%
\begin{equation}
\gamma =-\frac{1}{2}\left[ \pm \sqrt{2\varkappa c^{2}B_{0}^{2}+\left( \frac{%
ac\cos \theta }{2}\right) ^{2}}+\frac{ac\cos \theta }{2}\right] .
\label{gammaB0}
\end{equation}

Using the equation for $\Theta _{0}$ in (\ref{full}) one can estimate that 
\begin{equation*}
a=\left\vert \mathbf{\nabla }\Theta _{0}\right\vert \sim \varkappa
B_{0}^{2}L,
\end{equation*}%
with $L$ a characteristic length of the system, so that the $a$ terms can be
neglected in the expression (\ref{gammaB0}), to obtain%
\begin{equation}
\gamma =\pm \sqrt{\frac{\varkappa }{2}}B_{0}c,  \label{gamalinear}
\end{equation}%
much larger that in the case $\delta B_{z}=0$.

\subsection{\protect\bigskip Case $\mathbf{B}_{0}=0$}

For the case without zero order magnetic field, $\mathbf{B}_{0}=0$, and
perturbations $\delta \mathbf{E}$, $\delta \mathbf{B}$, $\delta \Theta \sim
\exp i\left( \mathbf{k}\cdot \mathbf{x}-\omega t\right) $ one has

\begin{eqnarray}
\mathbf{k}\cdot \delta \mathbf{E} &=&-\mathbf{k}\cdot \mathbf{E}_{0}\delta 
\mathbf{\Theta }+i\mathbf{\nabla }\Theta _{0}\cdot \delta \mathbf{E},  \notag
\\
\mathbf{k}\times \delta \mathbf{E} &=&\omega \delta \mathbf{B},\;\;  \notag
\\
\mathbf{k}\times \delta \mathbf{B} &=&-\frac{\omega }{c^{2}}\delta \mathbf{E}%
-\frac{\omega }{c^{2}}\mathbf{E}_{0}\delta \Theta +i\mathbf{\nabla \Theta }%
_{0}\times \delta \mathbf{B},  \notag \\
\left( k^{2}c^{2}-\omega ^{2}\right) \delta \Theta &=&-2\varkappa \mathbf{E}%
_{0}\cdot \delta \mathbf{E}.  \label{dTheta}
\end{eqnarray}

For propagation along the electric field, $\mathbf{E}_{0}\parallel \mathbf{k}
$, with $\mathbf{\nabla }\Theta _{0}$ $\perp \mathbf{k}$, one then has

\begin{eqnarray*}
\delta \Theta &=&\frac{\left( i\mathbf{\nabla }\Theta _{0}-\mathbf{k}\right) 
}{\mathbf{k}\cdot \mathbf{E}_{0}}\cdot \delta \mathbf{E}, \\
\left( \mathbf{k}-i\mathbf{\nabla }\Theta _{0}\right) \times \left( \mathbf{k%
}\times \delta \mathbf{E}\right) &=&-\frac{\omega ^{2}}{c^{2}}\delta \mathbf{%
E}+\frac{\omega ^{2}}{c^{2}}\mathbf{E}_{0}\frac{\left( \mathbf{k}-i\mathbf{%
\nabla }\Theta _{0}\right) }{\mathbf{k}\cdot \mathbf{E}_{0}}\cdot \delta 
\mathbf{E}.
\end{eqnarray*}
Again, it is easy to see that the last relation is satisfied, for arbitrary
longitudinal component $\delta E_{\shortparallel }$, only if $\omega
^{2}=k^{2}c^{2}$, with the standard plane EM wave relations for $\delta 
\mathbf{E}_{\perp }$ and $\delta \mathbf{B}$; while the longitudinal
component $\delta E_{\shortparallel }$ must be zero in order to satisfy (\ref
{dTheta}). In this way, the standard plane EM wave is the only solution in
this case.

For propagation perpendicular to the zero order field, one has now $\mathbf{B%
}_{0}=0$, $\mathbf{E}_{0}=E_{0}\mathbf{e}_{z}$, $\mathbf{\nabla }\Theta
_{0}=a\mathbf{e}_{x}$, $\mathbf{k}=k_{x}\mathbf{e}_{x}+k_{y}\mathbf{e}_{y}$,
and the general system (\ref{full}) can be reduced to%
\begin{eqnarray*}
\left( \omega ^{2}-k^{2}c^{2}\right) \delta B_{z}+iak_{x}c^{2}\delta B_{z}
&=&0, \\
\left( \omega ^{2}-k^{2}c^{2}\right) \delta B_{x}-iak_{y}c^{2}\delta B_{y}
&=&\frac{2\varkappa \omega ^{2}E_{0}^{2}k_{y}}{k^{2}\left( \omega
^{2}-k^{2}c^{2}\right) }\left( k_{x}\delta B_{y}-k_{y}\delta B_{x}\right) ,
\\
\left( \omega ^{2}-k^{2}c^{2}\right) \delta B_{y}+iak_{x}c^{2}\delta B_{y}
&=&-\frac{2\varkappa \omega ^{2}E_{0}^{2}k_{x}}{k^{2}\left( \omega
^{2}-k^{2}c^{2}\right) }\left( k_{x}\delta B_{y}-k_{y}\delta B_{x}\right) .
\end{eqnarray*}

If $\delta B_{z}\neq 0$, one has $\omega =kc+i\gamma $, with the same value (
\ref{small}) as in the case of $\mathbf{E}_{0}=0$, while if $\delta B_{z}=0$
one has the much larger value%
\begin{equation}
\gamma =\pm \sqrt{\frac{\varkappa }{2}}E_{0}.  \label{gammalinearE}
\end{equation}

\subsection{Experimental possibilities}

From the previous subsections it is seen that the more noticeable effects
are obtained for propagation perpendicular to the zero order fields, with
the appropriate polarization of the wave in each case. Moreover, comparing (%
\ref{gammalinearE}) with (\ref{gamalinear}) it is clear that, from a
practical point of view, magnetic fields are preferable. In any case, when
the beam traverses a length $\Delta L$, the relative variation of the
amplitude $A$ of any field is given by
\begin{equation}
\frac{\Delta A}{A}=\exp \left( \frac{\gamma \Delta L}{c}\right) .
\label{deltaa}
\end{equation}

For the case of a magnetic field of 1T, with $\delta B_{z}\neq 0$, one can
thus estimate from (\ref{chi}) and (\ref{gamalinear})%
\begin{equation*}
\frac{\gamma }{c}\sim 10^{-4}m^{-1}.
\end{equation*}

Although this effect is relatively large, there is the problem that both,
growing and decreasing modes are always present, so that a wave entering the
region with the static magnetic field results in a superposition of both
modes, and so the variation of amplitude of the growing mode cancels with
that of the decreasing mode at first order in $\gamma \Delta L/c$, and the
effect is only observable at second order, much hindering the experiment.

There is however a further possibility. It was argued in\cite%
{minotti} that, in order for the theory to be consistent with the lack of
strong gravitational effects due to the magnetic field of the Earth, the
non-linear terms in Eqs. (\ref{phi}) and (\ref{psi}) should come into play.
In this way, for the case of a static magnetic field outside its sources one
can write $\mathbf{B}=\mathbf{\nabla }\Psi $, with $\nabla ^{2}\Psi =0$, so
that equations (\ref{phi}) and (\ref{psi}) for the static case are%
\begin{eqnarray*}
\left( 2\omega _{0}+3\right) \nabla ^{2}\phi +\left. \frac{d\omega }{d\phi }%
\right\vert _{\phi _{0},\psi _{0}}\mathbf{\nabla }\phi \cdot \mathbf{\nabla }%
\phi  &\propto &B^{2}=\mathbf{\nabla }\Psi \cdot \mathbf{\nabla }\Psi , \\
\nabla ^{2}\psi +\mathbf{\nabla }\phi \cdot \mathbf{\nabla }\psi  &\propto
&B^{2}=\mathbf{\nabla }\Psi \cdot \mathbf{\nabla }\Psi ,
\end{eqnarray*}%
which have the exact solutions $\mathbf{\nabla }\phi \propto \mathbf{\nabla }%
\psi \propto \mathbf{\nabla }\Psi $, so that $\nabla ^{2}\phi =\nabla
^{2}\psi =0$, thus largely\ reducing the source of the gravitational force.
This solution for the case of the Earth's magnetic field is compatible with
the proposal in\cite{MLR}, in which the solution with $\nabla ^{2}\phi \neq 0
$ was used, if 
\begin{equation*}
\left. d\omega /d\phi \right\vert _{\phi _{0},\psi _{0}}\sim -\left( 2\omega
_{0}+3\right) .
\end{equation*}
With these considerations, the field $\Theta _{0}$ is simply given by%
\begin{equation*}
\mathbf{\nabla }\Theta _{0}=\lambda \mathbf{B}_{0},
\end{equation*}%
with $\lambda \simeq \sqrt{\chi }$. In this case the system (\ref{full}) can
be written as (with $\mathbf{E}_{0}=0$)%
\begin{eqnarray*}
\mathbf{\nabla }\cdot \delta \mathbf{E} &=&-\lambda \mathbf{B}_{0}\cdot
\delta \mathbf{E}, \\
\mathbf{\nabla }\times \delta \mathbf{E} &=&-\frac{\partial \delta \mathbf{B}%
}{\partial t},\;\;\;\mathbf{\nabla }\cdot \delta \mathbf{B}=0, \\
\mathbf{\nabla }\times \delta \mathbf{B} &=&\frac{1}{c^{2}}\frac{\partial
\delta \mathbf{E}}{\partial t}-\lambda \mathbf{B}_{0}\times \delta \mathbf{B}%
,
\end{eqnarray*}%
where the term with $\mathbf{\delta \Theta }$ can be neglected as it is
small, since from (\ref{full}) one can estimate that 
\begin{equation*}
\left\vert \frac{\mathbf{\nabla \delta \Theta }\times \mathbf{B}_{0}}{%
\mathbf{\nabla \Theta }_{0}\times \delta \mathbf{B}}\right\vert \sim \sqrt{%
\chi }B_{0}L,
\end{equation*}%
with $L$ a characteristic length of field variation.

Proceeding as before one has%
\begin{eqnarray}
\mathbf{k}\cdot \delta \mathbf{E} &=&i\lambda \mathbf{B}_{0}\cdot \delta 
\mathbf{E},  \notag \\
\mathbf{k}\times \left( \mathbf{k}\times \delta \mathbf{E}\right) &=&-\frac{%
\omega }{c^{2}}\delta \mathbf{E}+i\lambda \mathbf{B}_{0}\times \left( 
\mathbf{k}\times \delta \mathbf{E}\right) .  \notag
\end{eqnarray}

For propagation perpendicular to $\mathbf{B}_{0}$ one has%
\begin{eqnarray*}
\left( \frac{\omega ^{2}}{c^{2}}-k^{2}\right) \delta \mathbf{E}_{\perp }+%
\frac{\omega ^{2}}{kc^{2}}\delta E_{\parallel }\mathbf{k} &=&i\lambda
B_{0}\delta E_{\perp }\cos \alpha \,\mathbf{k}, \\
k\delta E_{\parallel } &=&i\lambda B_{0}\delta E_{\perp }\cos \alpha ,
\end{eqnarray*}%
where $\alpha $ is the angle between $\mathbf{B}_{0}$ and $\delta \mathbf{E}%
_{\perp }$. As a result the dispersion relation is that of a normal EM wave, 
$\omega =kc$, and the only anomalous effect is the presence of a small
longitudinal component of the electric field.

In the case of propagation along $\mathbf{B}_{0}$ one has%
\begin{eqnarray*}
\left( \frac{\omega ^{2}}{c^{2}}-k^{2}\right) \delta \mathbf{E}_{\perp }
&=&-i\lambda kB_{0}\delta \mathbf{E}_{\perp }, \\
k\delta E_{\parallel } &=&i\lambda B_{0}\delta E_{\parallel },
\end{eqnarray*}%
so that $E_{\parallel }=0$, and $\omega =kc+i\gamma $, with%
\begin{equation*}
\gamma =-\frac{\lambda B_{0}c}{2}.
\end{equation*}

This effect is similar in magnitude to that in (\ref{gamalinear}), but with
the advantage that only one sign is possible, so that there are not
coexisting growing and decaying modes, and the growth (or decay) could be
observed at first order.

\section{Table-top experiments with optical fibers}

Due to the relevance of first order effects in the propagation of EM waves,
it seems plausible that the use of simple and readily available nowadays
fiber optics would allow the verification of the theoretical
results of the previous section. Especially, the last result of subsection
2.3 shows a direct method for measuring the additional amplitude change caused by
the propagation of a single mode inside an ordinary polymer fiber. Given the
significance of separating between alternative extentions of general
relativistic theories for modern cosmology we propose that such an
experiment is of great importance due to its simplicity.

Specifically, our proposal is to get a sufficiently large fiber appropriately coiled which, with existing materials, can be made to easily reach a km of total
distance for the propagating mode. By taking the
logarithm of Eq (\ref{deltaa}) as $10log_{10}(\frac{\Delta A}{A})$, for the
amplitude variation to be in dB units, we see that a magnetic field of 2 T
would result in an amplitude variation of 1 dB in 1 km distance. It is
possible to reduce such a distance by an order of magnitude only through a
large magnetic field of about 10 T or more which can be produced in current
NMR devices while use of superconducting elements could reach even higher values.
Actually, recent reports from the NHMFL at Los Alamos claim a 100 T machine
is already operational \cite{bigmagnet}. At the moment we will only assume
the strongest existing rare earth magnets like Boron - Neodymium for a
tabletop experiment where sufficiently high accuracy power meters are
available. The central idea is to detect the difference between measurement on the fiber coil, with and without the B field.
Present day power meters have an accuracy noise threshold of about $0.1$ dB.
Fortunately, existing manufacturers may be able to provide bobbins totalling
25 km of fiber or more so that a measurement of 10 - 20 dB of additional
amplitude variations is in principle possible.

\begin{figure}[h]
\includegraphics[width=1.0\textwidth]{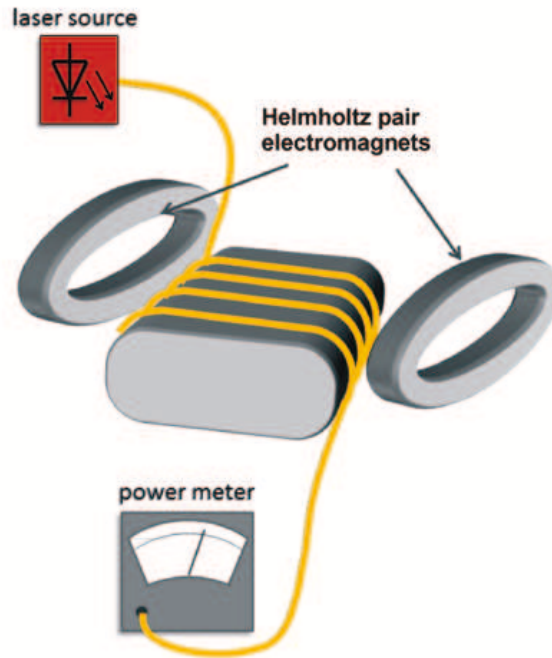}
\caption{\label{fig1}Proposed configuration of the optical fiber and Helmholtz coils.}
\end{figure}

With respect to the fiber coiling process, one has to take into
account that any angles introduced to the fiber material introduce additional
attenuation to any propagating mode. Technical data for existing fiber
materials suggest that there should be a certain curvature with angles small enough
not to cause severe damping during normal propagation. This can be achieved with a flattened coil frame like
the one shown in Fig.\ref{fig1}. 

As we are not interested in all the engineering details of an actual
experiment we only emphasize the main points where care must be taken using some simplified
configurations. In the flattened fiber coil of Fig.\ref{fig1} care must be taken so that on the upper path
the fiber is parallel to the direction of the external
magnetic field. The return path, though, must be outside the region of influence or else the 
amplitude variation effect will be cancelled and no difference will be measured. For this reason we also
made the flattened electromagnets shown in such a way that the homogenized flux of the applied B 
field  will only affect the upper part of the fiber's path.
It is also possible to make up an homogeneous magnetic field using Neodymium magnets in a special configuration known as a cylindrical "Halbach Array" \cite{Halbach}. Such arrays have been in use for a long time in magnetic trains, very fast brushless motors and similar electrical engineering applications.
In such a case, the upper or lower part of the flattened fiber coil should be put inside the region of homogeneous B flux of a Halbach cylinder.

\section{Conclusions}

We have here reported for the first time some new results on the possible gravitational influence on
classical EM fields in scalar-tensor extentions of General Relativity. We also used the linearized version 
of the perturbed Maxwell equations to analyse the propagation of ordinary modes. The analysis led us to 
conclude the possibility of easy, low cost experiments with fiber optics that would allow the verification 
of the said theories.
We believe that the present state of cosmology with the recurring acute problems of inflation and initial 
conditions, the CMB anisotropy as well as the dark matter and dark energy, fully justifies the continuation
of the present research in more areas where evidence can be accumulated experimentally.


\begin{thebibliography}{99}
\bibitem{fujiibook} Fujii Y., Maeda K. 2003 \textit{The Scalar-Tensor Theory
of Gravitation} {Cambridge University Press, Cambridge U. K.}

\bibitem{chavineau} Chavineau, B. 2007 \textit{Phys. Rev. D} \textbf{76}
104023.

\bibitem{bransdicke} Brans C., Dicke R. H. 1961 \textit{Phys. Rev.} \textbf{%
124} 925.

\bibitem{fifthforce} Fishbach E. and Talmadge C. 1992 \textit{Nature} 
\textbf{356}, 207.

\bibitem{frt} De Felice A. and Tsujikawa S. 2010 \textit{Living Rev.
Relativity} \textbf{13} 3.

\bibitem{bekenstein} Bekenstein J. D. 1982 \textit{Phys. Rev. D} \textbf{25}
1527.

\bibitem{bertotti} Bertotti B., Iess L., Tortora P. 2003 \textit{Nature} 
\textbf{425} 374.

\bibitem{MLR} Mbelek J. P. and Lachi\`{e}ze-Rey M. 2002, \textit{Grav. \&
Cosmol.} \textbf{8} 331.

\bibitem{mbelek2003} Mbelek J. P. and Lachi\`{e}ze-Rey M. 2003, \textit{%
Astron. Astrophys.} \textbf{397} 803.

\bibitem{mbelek2004} Mbelek J. P. 2004, \textit{Astron. Astrophys.} \textbf{%
424} 761.

\bibitem{minotti} Minotti F. O. 2013, \textit{Grav. \& Cosmol.} (in press)
(arXiv:1302.5690 [gr-qc]).

\bibitem{juan} Yang J., Wang Y.-Q., Li P.-F., Wang Y., Wang Y.-M., Ma Y.-J.
2012, \textit{Acta Phys. Sin.} \textbf{61} 110301.

\bibitem{bigmagnet} J. L. Bacon \textit{et al}, 2002, IEEE Trans. App. Supercond. \textbf{12}(1) 695.

\bibitem{Halbach} K. Halbach, 1980, Nuc. Inst. Meth. \textbf{169}(1) 1.

\bibitem{Plancksat} A. Ijjas \textit{et al}, 2013, Phys. Let. B, (In Press)
\end{thebibliography}
\end{document}